# L-Mode and Inter-ELM Divertor Particle and Heat Flux Width Scaling on MAST

J. R. Harrison*[a], G. M. Fishpool[a], A. Kirk[a]

[a]*EURATOM/CCFE Fusion Association, Culham Science Centre, Abingdon, OX14 3DB, UK.*

**Abstract**

The distribution of particles and power to plasma-facing components is of key importance in the design of next-generation fusion devices. Power and particle decay lengths have been measured in a number of MAST L-mode and H-mode discharges in order to determine their parametric dependencies, by fitting power and particle flux profiles measured by divertor Langmuir probes, to a convolution of an exponential decay and a Gaussian function. In all discharges analysed, it is found that exponential decay lengths mapped to the midplane are mostly dependent on separatrix electron density ($n_{e,sep}^{0.65\pm0.15}$ L-mode, $n_{e,sep}^{0.76\pm0.19}$ H-mode) and plasma current ($I_p^{-0.36\pm0.11}$ L-mode, $I_p^{-1.05\pm0.18}$ H-mode) (or parallel connection length). The widths of the convolved Gaussian functions have been used to derive an approximate diffusion coefficient, which is found to vary from $1 m^2/s$ to $7 m^2/s$, and is systematically lower in H-mode compared with L-mode.

*\*Corresponding Author Address*: EURATOM/CCFE Fusion Association, Culham Science Centre, Abingdon, OX14 3DB, UK

*\*Corresponding Author E-mail*: james.harrison@ccfe.ac.uk

*Presenting Author*: James Harrison

*Presenting Author E-mail*: james.harrison@ccfe.ac.uk

## 1. Introduction

The steady-state distribution of power and particles to plasma-facing components is of key importance when assessing the feasibility of divertor solutions in next-generation fusion devices. There are significant uncertainties concerning the transport mechanisms that govern this distribution, impeding efforts to calculate divertor heat and particle loads from first principles. Instead, estimates of heat flux profiles in future devices rely upon extrapolating divertor heat flux characteristics measured in existing devices. Furthermore, the design of baffling structures to increase the divertor neutral compression requires estimates of the particle flux decay lengths. This work follows from a study carried out on JET and ASDEX Upgrade [1] to characterise the divertor heat flux profile by an exponential decay length to account for the decay of $n_e$, $T_e$, upstream, convolved with a Gaussian profile to account for diffusive processes below the x-point. The total profile shape, $q(\bar{s})$, is given by:

$$q(\bar{s}) = \frac{q_0}{2} \exp\left[\left(\frac{S}{2\lambda_q f_x}\right)^2 - \frac{\bar{s}}{\lambda_q f_x}\right] \mathrm{erfc}\left(\frac{S}{2\lambda_q f_x} - \frac{\bar{s}}{S}\right) + q_{BG} \quad (1)$$

where $\bar{s} = s - s_0$, the difference between a given target radial co-ordinate and the strike location (m), S is the width of the Gaussian function (m), $\lambda_q$ is the exponential decay length as measured at the divertor target (m), $f_x$ is the flux expansion from upstream to the divertor and $q_{BG}$ is a background heat or particle flux density. A database has been compiled of 215 L-mode and 95 H-mode lower outer divertor heat and particle flux profiles as measured by flush-mounted single Langmuir probes embedded in divertor target plates. All divertor strike points are attached in the data analysed. The data were analysed from stationary periods of MAST discharges free of transient MHD activity or ELMs to improve the quality of the data collected. The database spans an operational space summarised in Table 1 where $I_p$ is plasma current, $B_\varphi$ is the toroidal field at the magnetic axis, determined by magnetic diagnostics and the EFIT equilibrium reconstruction code. $P_{SOL}$ is the power crossing the separatrix, the total Ohmic and auxiliary heating power minus radiated power as measured by bolometers and the rate of change of stored magnetic energy from EFIT. $n_{e,sep}$ is the upstream separatrix electron density, which is determined using a 130 point Thomson scattering diagnostic, which measures radial profiles of electron temperature and density with ~10mm spatial resolution. Determination of the separatrix electron density is subject to significant uncertainties as the location of the separatrix can be estimated to within ±10mm due to uncertainties in the calculated equilibrium. $L_{\parallel}$ is the parallel connection length, the distance along a field line from 10mm radially outboard of the separatrix at the midplane, to lower outer divertor target, calculated using magnetic field information from EFIT and a field line following code.

## 2. Particle Flux Profiles

Divertor particle flux profiles derived from measurements of ion saturation current, $j_{sat}$, as measured by the target probes, as $\Gamma_i(s) = e\, j_{sat}(s)$, where $\Gamma_i$ is the ion flux incident to the probe (ions/m$^2$/s) and e is the elementary charge. Moreover, divertor power flux profiles are calculated by $P_{tar}(s) = \gamma j_{sat}(s) T_e(s)$ where $\gamma$ is the sheath heat transmission coefficient, taken to be 7 in this study, which assumes $T_i = T_e$. Comparison between particle and power decay lengths is carried out by fitting profiles of $\Gamma_i(s)$ and $P_{tar}(s)$ to $q(\bar{s})$ (equation 1) to determine $\lambda_q f_x$. Figure 1 shows that the decay lengths of particle and power fluxes, $\lambda_{jsat}$ and $\lambda_q$, are comparable over a wide range of discharges, indicating that $T_e$ decays weakly with radius [2], (i.e. $\lambda_{ne} < \lambda_{Te}$), therefore the power decay length is dominated by the density decay length. Similar behaviour has been observed on ASDEX Upgrade [3].

## 3. Power Flux Profiles

Two commonly applied measures of the decay length of power flux profiles are the midplane exponential decay length, $\lambda_q$, and the integral width, defined as [1]:

$$\lambda_{int} = \frac{\int (q(s) - q_{BG})ds}{\max(q(s))f_x} \qquad (2)$$

where the region of integration is taken to cover the heat flux profile over ±10cm from the profile peak. Scaling of both $\lambda_q$ and $\lambda_{int}$ has been performed in this study as $\lambda_q$ describes the exponential decay of heat flux profiles in the near-SOL and is governed only by the physics affecting the near-SOL. $\lambda_{int}$ is an engineering parameter used to relate the peak to the integrated heat flux to allow straightforward estimation of the divertor peak heat flux in future devices, however, it is affected by the physics governing both the near and far SOL. Linear regression of $\lambda_q$ and $\lambda_{int}$ for both L-mode and H-mode discharges (Figure 2) shows that:

$$\lambda_{int,L}(mm) = (8.6 \pm 0.6) + (0.96 \pm 0.05)\lambda_q(mm) \qquad (3)$$

$$\lambda_{int,H}(mm) = (3.7 \pm 0.5) + (1.41 \pm 0.06)\lambda_q(mm) \qquad (4)$$

the increased constant term in the L-mode scaling reflects an increase in the width of the power flux profile in the private flux region, via the S term in equation 1, thereby increasing $\lambda_{int}$, and is unlikely to be linked to changes in divertor conditions such as enhanced radiative losses observed during detachment. The regression analysis of the H-mode data shows larger coefficients than observed on JET or ASDEX Upgrade [1], although $\lambda_{int} \approx 2 \times \lambda_q$ for the MAST data, over most of the data range, in agreement with findings from other machines.

## 4. L-Mode Heat Flux Scaling

Regression analysis of $\lambda_q$ in L-mode discharges with respect to $P_{SOL}$, $I_p$, $n_{e,sep}$ and $B_\varphi$ over the range of operating parameters shown in table 1 yields the following scaling expression:

$$\lambda_q(m) = (1.2 \times 10^{-2}) \times P_{SOL}^{0.14 \pm 0.09}(MW) I_P^{-0.36 \pm 0.11}(MA) n_{e,sep}^{0.65 \pm 0.15}(10^{19} m^{-3}) B_\phi^{-0.09 \pm 0.25}(T) \quad (5)$$

The above expression exhibits a weaker density scaling than has been observed previously on MAST [4] $\lambda_q \propto n_e^{-1.52 \pm 0.16}$, although the discrepancy may be due the use of the separatrix density in this study, as opposed to the line-averaged density used in the previous study. The use of the separatrix density in this study removes any sensitivity to the variation in density profile due to density peaking or particle fuelling sources crossing the diagnostic sightline. If the power decay length scaling is based instead on line average density, $\overline{n_e}$, it is found that $\lambda_q \propto n_e^{-1.27 \pm 0.25}$ albeit with a decreased quality of fit to the complete dataset. The uncertainty in the power to which $n_{e,sep}$ is raised is dominated by uncertainties in the location of the separatrix. If the database is fit with a power law that replaces $I_p$ and $n_{e,sep}$ with the fraction of the Greenwald density limit ($f_{GW} = \dfrac{\overline{n_e} \pi a^2}{I_p}$) in (5), it is found that $\lambda_q \propto f_{GW}^{0.30 \pm 0.17}$ and the powers to which $P_{SOL}$ and $B_\varphi$ is raised is unaffected. The scaling with respect to toroidal field strength has considerable uncertainties due to the limited range over which $B_\varphi$ was varied in the available dataset. The presence of a density dependence is in contrast to recent results from JET and ASDEX Upgrade [1], but in agreement with inter-machine studies [5], emphasising the difficulties in determining the operating parameters to include in such scaling relations.

To calculate a scaling law more appropriate to future long-legged divertors, such as the Super-X of MAST-Upgrade[6], a regression analysis of $\lambda_q$ with respect to $L_\parallel$ was carried out (Figure 3). This results in the following expression for $\lambda_q$:

$$\lambda_q(m) = (6.7 \times 10^{-3}) \times P_{SOL}^{0.1 \pm 0.13}(MW) L_{\parallel}^{0.34 \pm 0.18}(m) n_{e,sep}^{0.69 \pm 0.13}(10^{19} m^{-3}) B_\phi^{0.03 \pm 0.23}(T) \qquad (6)$$

Application of the above expression to predict $\lambda_q$ in MAST-Upgrade, taking $P_{SOL} \sim 11$MW, $n_{e,sep} \sim 1 \times 10^{19}$m$^{-3}$, $B_\varphi = 0.8$T, $L_{\parallel} = 10$m (conventional), 20m (Super-X), yields $\lambda_q \sim 18$mm (conventional), ~23mm (Super-X). Conversely, application of (5) to determine $\lambda_q$ in MAST-Upgrade, taking $I_p = 1.3$MA yields a prediction of $\lambda_q = 13$mm, which is independent of divertor magnetic configuration.

## 5. H-Mode Heat Flux Scaling

Regression analysis of the variation in $\lambda_q$ in H-mode discharges was carried out with respect to $P_{SOL}$, $I_p$, $n_{e,sep}$ and $B_\varphi$ over a range of operational parameters shown in table 1. The data was found to obey the following expression:

$$\lambda_q(m) = (4.5 \times 10^{-3}) \times P_{SOL}^{0.11 \pm 0.18}(MW) I_P^{-1.05 \pm 0.18}(MA) n_{e,sep}^{0.76 \pm 0.19}(10^{19} m^{-3}) B_\phi^{-0.07 \pm 0.37}(T) \qquad (7)$$

Both L-mode and H-mode scalings exhibit weak sensitivity to $P_{SOL}$, in agreement with findings from NSTX [7], DIII-D [8], JET and ASDEX Upgrade [1], although weaker than was found in inter-machine scaling studies [5]. The scaling with respect to $B_\varphi$ is subject to considerable uncertainties due to the limited range over which the toroidal field was varied. The range of $n_{e,sep}$ spanned in the database is limited to the region around the density required to minimise the L-H transition power [9]. The uncertainty in absolute value of $n_{e,sep}$ is due to random errors in the measurement and in determining the location of the separatrix. As before, repeating this analysis with $I_p$ and $n_{e,sep}$ replaced by the fraction of the Greenwald density limit, $f_{GW}$, results in $\lambda_q \propto f_{GW}^{0.63 \pm 0.16}$ with little variation in the powers to which $P_{SOL}$ and $B_\varphi$ is raised in (7), compared to their uncertainties. $\lambda_q$ is found to have greatest sensitivity to $I_p$, in common with DIII-D [8], NSTX [7] and a heuristic model based on drifts in the SOL [10], although the $I_p$ scaling is weaker than observed on NSTX, $I_p^{-1.6\pm0.1}$, within stated uncertainties. Scaling with respect to $L_\parallel$ (Figure 4) yields the following expression for $\lambda_q$:

$$\lambda_q(m) = (7.3\times10^{-4})P_{SOL}^{0.19\pm0.15}(\text{MW})L_\parallel^{1.01\pm0.43}(\text{m})n_{e,sep}^{0.65\pm0.21}(10^{19}\text{m}^{-3})B_\phi^{-0.10\pm0.33}(\text{T}) \qquad (8)$$

indicating a favourable scaling for divertor designs that maximise $L_\parallel$, to a greater degree in H-mode compared with the comparable L-mode scaling. Application of (8) to predict $\lambda_q$ for MAST-Upgrade conventional and Super-X configurations, assuming the same machine parameters as used in the L-mode prediction, yields $\lambda_q \sim$ 11mm (conventional), 23mm (Super-X). Both L-mode and H-mode scalings predict similar $\lambda_q$ for Super-X equilibria, which may be a consequence of extrapolating, in terms of $L_\parallel$, which was varied over a limited range, approximately a factor of 2 beyond the data set used in this study. Therefore, the stated extrapolations are subject to considerable uncertainties. Application of (7) to predict $\lambda_q$ in MAST-Upgrade yields $\lambda_q$ = 4.5mm, suggesting that whether $I_p$ or $L_\parallel$ is a dominant scaling parameter has a significant effect on the projected power decay length in future devices.

## 6. Estimating Transport Coefficients

The description of the divertor heat flux profile in equation (1) can be calculated as the convolution of an exponential function of decay length $\lambda_q$ and a Gaussian function of the form $q_0.\exp[-s^2/\sigma^2]$, where $\sigma = S/(2\times\ln(2))$. Conversely, the analytic solution of the 1-D diffusion equation for the problem of propagation of heat given fixed upstream conditions is a Gaussian function whose $\sigma = (2\times D\tau)^{0.5}$, where D is a diffusion coefficient and $\tau$ is the timescale over which diffusive transport occurs. Therefore, using the data used to calculate $\lambda_q$, it is possible to estimate the cross-field (particle) diffusivity coefficient for a given profile, assuming the transport timescale $\tau$ is known. In the analysis presented, it is assumed that parallel temperature gradients are sufficiently small that the upstream temperature can be taken as an estimate for $T_e$ at the x-point, and that $T_e = T_i$, as there are no data concerning $T_e$ or $T_i$ at the x-point on MAST. An estimate of $\tau$ is given by $L_\parallel(xpt)/c_s$, where $L_\parallel(xpt)$ is the parallel connection length between the x-point and the divertor strike point, and $c_s$ is the ion acoustic velocity. The assumption is somewhat justified as the SOL collisionality (ratio of the electron-electron mean free path and parallel connection length from upstream to target) in the discharges of interest are in the region of 5-25 in L-mode and 2-10 in H-mode data. Comparison of $S^2$ with $\tau$ (Figure 5) shows that the Gaussian broadening, which determines the radial extent of the private flux region and the far-SOL, is a function of confinement regime. In H-mode discharges, it is noted that single-null diverted (SND) discharges have a systematically higher value of S than double-null diverted (DND) discharges in H-mode. A possible explanation for this dependency is that the in:out divertor heat load asymmetry is 1:30 in DND configuration and up to 1:3 in SND configuration [11], and therefore it is likely that SOL transport is sensitive to this change in magnetic configuration. The subject of future work will be to investigate the sensitivity of S to changes in SOL transport to determine if this is indeed the case. The L-mode data presented are in DND configuration, so no such variation was observed in these data. Comparison of the data with values of different diffusion coefficients suggests that, within uncertainties, the DND L-mode and H-mode data are bounded within $1m^2/s \leq D \leq 5m^2/s$, which is consistent with modelling of the MAST SOL [12], and that SND H-mode data seem to have a greater diffusivity, which may be explained in terms of changes in the in:out asymmetry described previously.

**Acknowledgements**

This work was funded by the RCUK Energy Programme under grant EP/I501045 and the European Communities under the contract of Association between EURATOM and CCFE. The views and opinions expressed herein do not necessarily reflect those of the European Commission.

**Figure Captions**

Figure 1: Comparison of mid-plane exponential decay lengths of ion saturation current, $\lambda_{jsat}$, and power incident to the probes, $\lambda_q$.

Figure 2: Midplane exponential decay lengths and integral widths in L-mode and H-mode.

Figure 3: Measured L-mode midplane power exponential decay length, $\lambda_q$, compared with scaling law prediction in terms of $P_{SOL}$, $L_{\parallel}$, $n_{e,sep}$, $B_{\varphi}$ (equation 6).

Figure 4: Measured H-mode midplane power exponential decay length, $\lambda_q$, compared with scaling law prediction in terms of $P_{SOL}$, $L_{\parallel}$, $n_{e,sep}$, $B_{\varphi}$ (equation 8).

Figure 5: Profile Gaussian width, S, and estimated SOL transport time $\tau$, and estimated SOL particle diffusivity for L-mode and H-mode discharges.

**Table Captions**

Table 1: Range of machine operating parameters from pulses where L-mode (L) and H-mode (H) data were taken.

# Figures

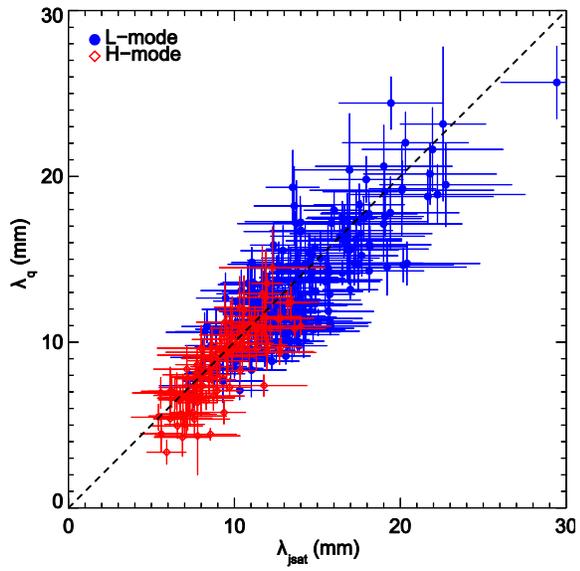

**Figure 1**

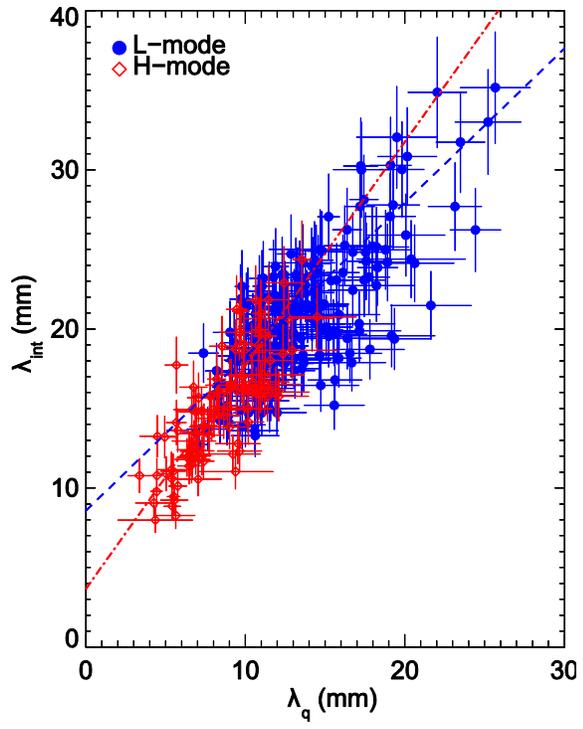

Figure 2

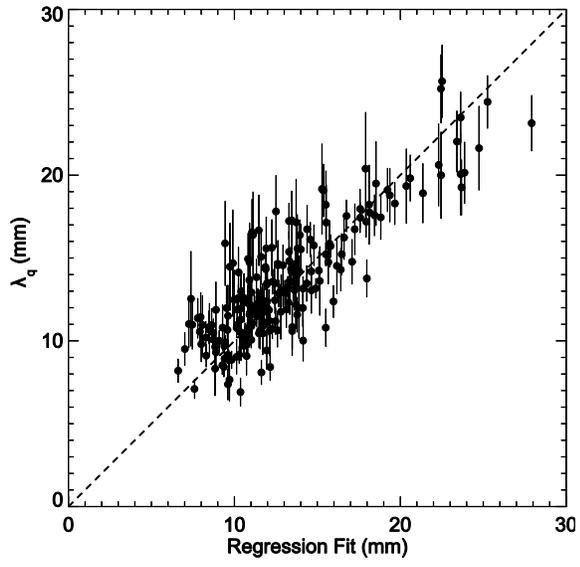

**Figure 3**

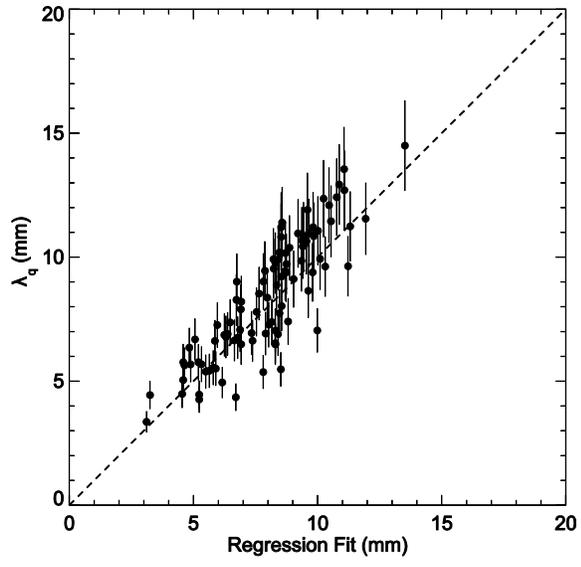

**Figure 4**

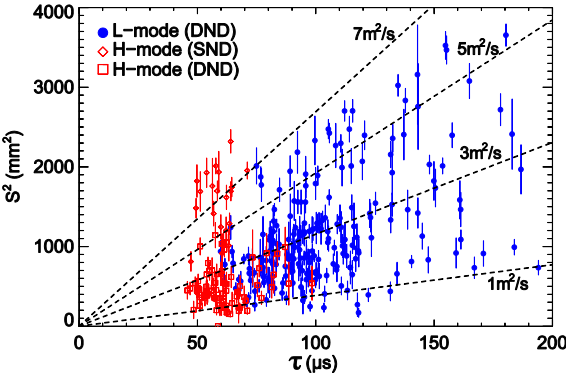

**Figure 5**

| Parameter | Min (L) | Max (L) | Min (H) | Max (H) |
|---|---|---|---|---|
| $P_{SOL}$ (MW) | 0.5 | 2 | 0.5 | 3.5 |
| $I_p$ (kA) | 400 | 900 | 550 | 900 |
| $n_{e,sep}$ ($10^{19}$ m$^{-3}$) | 0.4 | 2 | 1 | 2 |
| $B_\varphi$ (T) | -0.45 | -0.56 | -0.45 | -0.53 |
| $L_\parallel$ (m) | 8 | 12 | 8 | 12 |

**Table 1**